\def\be{\begin{equation}}
\def\ee{\end{equation}}
\def\ba{\begin{array}}
\def\ea{\end{array}}

\def\Nb{{I\!\! N}}

\def\Cb{\ \hbox{\vrule width 0.6pt height 6pt depth 0pt
		      \hskip -3.2 pt} C}
\documentstyle[12pt]{article}
\begin{document}
\parskip=4pt
\parindent=18pt
\baselineskip=22pt
\setcounter{page}{1}
\centerline{\Large\bf Teleportation of general finite dimensional}
\vspace{2ex}
\centerline{\Large\bf quantum systems}
\vspace{6ex}
\begin{center}
{\large  Sergio Albeverio}\footnote {\raggedright SFB 256; SFB  237; BiBoS;
CERFIM (Locarno); Acc.Arch., USI (Mendrisio)\\
~~~~~e-mail: albeverio@uni-bonn.de}~~~ and ~~~
{\large  Shao-Ming Fei}\footnote{\raggedright Institute of Physics,
Chinese Academy of Science, Beijing\\
~~~~~e-mail: fei@uni-bonn.de}
\end{center}
\begin{center}
Institut f\"ur Angewandte Mathematik, Universit\"at Bonn,
D-53115 Bonn\\
Fakult\"at f\"ur Mathematik, Ruhr-Universit\"at Bochum
D-44780 Bochum\\
\end{center}
\vskip 1 true cm
\parindent=18pt
\parskip=6pt
\begin{center}
\begin{minipage}{5in}
\vspace{3ex}
\centerline{\large Abstract}
\vspace{4ex}
Teleportation of finite dimensional quantum states
by a non-local entangled state is studied. 
For a generally given entangled state,
an explicit equation that governs the teleportation
is presented. Detailed examples and the roles played
by the dimensions of the Hilbert spaces related to the sender, receiver
and the auxiliary space are discussed.

\bigskip
\medskip
\bigskip
\medskip


AMS classification codes: 81P68 68P30

\end{minipage}
\end{center}

\newpage

The discovery of quantum teleportation protocols belongs to the most
important results of quantum information theory.
Quantum teleportation has been introduced in \cite{Bennett93}
and discussed by a number of authors for both spin-$\frac{1}{2}$
states and arbitrary quantum states, see e.g. [2-11].
By means of a classical and a distributed quantum communication channel,
realized by a non-local entangled state chosen in a
special way (e.g., an EPR-pair of particles),
the teleportation process allows to transmit an unknown quantum state
from a sender (traditionally) named
``Alice" to a receiver ``Bob" that are spatially separated.
For teleportation of $N$-dimensional quantum states, the
teleportation problem has been discussed in \cite{luigi}
in the case where
the dimensions of the Hilbert spaces associated with the sender,
receiver and the auxiliary space are all equal to $N=2^{m}$,
for a given $m\in\Nb$. The relations among
quantum teleportation, dense coding,
orthonormal bases of maximally entangled vectors
and unitary operators with respect to the 
Hilbert-Schmidt scalar product, and depolarizing operations
are investigated in \cite{werner}.

The details of protocols for teleportation vary with
the shared entangled state and joint measurements at Alice or Bob.
In this note we study the general properties of 
teleportation for finite dimensional quantum states without the 
assumption on equality for the dimensions of the Hilbert spaces involved.
We give a teleportation protocol for generally given entangled states
and a constraint equation that governs the teleportation.
The solutions of the constraint equation give
the unitary transformations of teleportation protocols.
Detailed examples and the roles played
by the dimensions of the Hilbert spaces are discussed.

Let $H_1$ be a Hilbert space with dimensions $N_1$.
Let $e_i$, $i=1,...,N_1$, be an orthogonal basis in $H_1$, so that $e_i$ 
is an $N_1$-dimensional column vector with entry $1$ for the $i$-th
component and $0$ for the other components. Alice has a general
quantum state on the Hilbert space $H_1$ of the form
\be\label{psi0}
\Psi_0=\left(\ba{c}\alpha_1\\ \vdots\\ \alpha_{N_1} 
\ea \right)=\sum_{i=1}^{N_1}\alpha_i e_i,
\ee
where $\alpha_i\in \Cb$, $\sum_{i=1}^{N_1}\vert\alpha_i\vert^2=1$.
For convenience we will call an arbitrary dimensional quantum state also
a qubit in the following.

Let $H_2$ and $H_3$ be auxiliary Hilbert spaces attached to Alice and Bob,
with dimensions $N_2$ and $N_3$ respectively. To send the state
$\Psi_0$ to Bob's hand, it is necessary that $N_3\geq N_1$.
Let $f_i$ (resp. $g_j$), $i=1,...,N_2$ (resp. $j=1,...,N_3$), be the
corresponding
orthogonal basis vector of the Hilbert space $H_2$ (resp. $H_3$).  
A generally entangled state of two qubits in the Hilbert spaces $H_2$
and $H_3$ is of the form
\be\label{psi1}
\Psi_1=\sum_{i=1}^{N_2}\sum_{j=1}^{N_3}a_{ij} f_i \otimes g_j
\ee
for some (normalized) complex coefficients $a_{ij}\in\Cb$. The degree of 
entanglement depends on the $a_{ij}$, $i=1,...,N_2$, $j=1,...,N_3$.
In the following we take $N_3=N_1$.

The initial state Alice and Bob have is then given by
\be\label{psi}
\Psi_0 \otimes \Psi_1=\sum_{i,k=1}^{N_1}\sum_{j=1}^{N_2}
\alpha_i a_{jk}e_i \otimes f_j \otimes g_k
~~~~~~\in H_1\otimes H_2\otimes H_3\,.
\ee
Alice has the first and the second qubits and Bob has the third one.
To transform the state of Bob's qubit to be $\Psi_0$ (given by
(\ref{psi0})), one has to do some
unitary transformation $U$ and measurements. The effect of these
operations together is called quantum ``teleportation".
Let $U$  be the unitary transformation
acting on the tensor product of 
two quantum states in the Hilbert spaces $H_1$ and $H_2$ such that
\be\label{u}
U (e_i \otimes f_j)=\sum_{s=1}^{N_1}
\sum_{t=1}^{N_2}b_{ijst} e_s \otimes f_t\,,
\ee
with $\displaystyle\sum_{s=1}^{N_1}\sum_{t=1}^{N_2}\vert b_{ijst}\vert^2=1$,
$\forall i=1,...,N_1$, $j=1,...,N_2$.

\medskip
{\sf [Theorem]}. If $b_{ijst}$ satisfies the following relation
\be\label{cond}
\sum_{i=1}^{N_1}\sum_{j=1}^{N_2}\alpha_i a_{jk} b_{ijst}=
\frac{1}{\sqrt{N_1N_2}}\alpha_{k-t+1}c_{s\,k-t+1\,t}
\ee
for some $c_{ijk}\in\Cb$ such that $c_{ijk}c_{ijk}^\ast=1$,
$U$ is the unitary transformation that fulfills the quantum teleportation.

{\sf [Proof]}. From (\ref{u}) and (\ref{cond}) we have, with $\Psi_0$
resp. $\Psi_1$ as in (\ref{psi0}) resp. (\ref{psi1}):
$$\ba{rcl}
(U\otimes 1)(\Psi_0 \otimes \Psi_1)\equiv\Phi&=&
\displaystyle\sum_{i,s,k=1}^{N_1}\sum_{j,t=1}^{N_2}
\alpha_i a_{jk}b_{ijst}\,e_s \otimes f_t \otimes g_k\\[4mm]
&=&\displaystyle\frac{1}{\sqrt{N_1N_2}}
\sum_{s,k=1}^{N_1}\sum_{t=1}^{N_2}
\alpha_{k-t+1} c_{s\,k-t+1\, t}
\,e_s \otimes f_t \otimes g_k\\[4mm]
&=&\displaystyle\frac{1}{\sqrt{N_1N_2}}
\sum_{i,j=1}^{N_1}\sum_{k=1}^{N_2}
c_{ijk}\alpha_j\, e_i \otimes f_k \otimes g_{k+j-1},
\ea
$$
where the sub indices of $e$, $f$ and $g$ are understood to be taken
modulo by $N_1$, $N_2$ and $N_3$ respectively. 

Now Alice measures her two qubits in the state $\Phi\in H_1\otimes H_2$. 
If $e_i \otimes f_k$ is the state obtained after the
measurement, i.e.,
$$
\Phi\to e_i \otimes f_k \otimes
\left(\sum_{j=1}^{N_1}c_{ijk}\alpha_j g_{k+j-1}\right),
$$
then in order to recover the original state $\Psi_0$,
the unitary operator that Bob should use to act on his qubit is
\be\label{oik}
O_{ik}=P_k C_{ik},~~~~~~~i=1,...,N_1,~~~k=1,...,N_2,
\ee
where $P_k$ is the $(k-1)$-th power of the permutation operator,
$P_k=\Pi^{k-1}$,
$$
\Pi=\left(\ba{ccccc}
&&&&1\\
1&&&&\\
&1&&&\\
&&\ddots &&\\
&&&1&\ea
\right)
$$
and $C_{ik}=diag(c_{i1k}^\ast,c_{i2k}^\ast,...,c_{iN_1 k}^\ast)$.
After this transformation, one gets $\Phi\to e_i\otimes
e_k\otimes\Psi_0$ and the state $\Psi_0$ given by (\ref{psi0}) is
teleported from Alice to Bob. \hfill $\rule{3mm}{3mm}$

Therefore whenever an entangled state in the sense of (\ref{psi1})
is given, i.e. the $a_{ij}$ are given, if there 
are solutions of $b_{ijst}$ to equation (\ref{cond}), we have a unitary 
transformation $U$ that fulfills the teleportation. The condition (\ref{cond})
can be rewritten as 
\be\label{cond1}
\sqrt{N_1 N_2}\sum_{i=1}^{N_2}a_{i\,t+j-1}b_{jist}=c_{sjt}\,.
\ee
The unitary transformation (\ref{u}) given by the
quantities $b_{jist}$ used in our teleportation protocol
depends on the initially given entangled state (\ref{psi1}) and the
dimensions of the Hilbert spaces $H_1$, $H_2$, $H_3$.

For general $N\equiv N_1=N_2=N_3$, if we take
\be\label{aij}
a_{ij}=\frac{\delta_{ij}}{\sqrt{N}},
\ee
the entangled state (\ref{psi1})
is given by
$$
\Psi_1=\frac{1}{\sqrt{N}}\sum_{i=1}^{N}f_i \otimes g_i.
$$
From equation (\ref{cond1}), we
obtain the unitary transformation (\ref{u}) used in the
teleportation protocol with
\be\label{bb}
b_{i\,t+i-1\, st}=\frac{c_{sit}}{\sqrt{N}},
\ee
with $c_{sit}$ as in (\ref{cond}),
the other coefficients $b$ in (\ref{u}) being zero.
It is easily checked that the transformation (\ref{u}) given by (\ref{bb})
is a unitary one. The teleportation is accomplished by applying the unitary
operation (\ref{oik}) according to the result of Bob's measurement.
For some particular values of the coefficints $c_{sit}$, this 
result concides with the
one in \cite{BBPSSW}.

According to the Schmidt decomposition, in this case the entangled state
(\ref{psi1}) on Hilbert spaces $H_2$ and $H_3$ can be always written as
$$
\sum_{i=1}^{N} \sqrt{\lambda_i} f_i \otimes g_i
$$
in suitable basis, where $\lambda_i\ge 0$, $\sum_{i=1}^N\lambda_i=1$.
That is, $a_{ij}=\sqrt{\lambda_i}\delta_{ij}$. Substituting it into
equation (\ref{cond1}), we have
$$
\sqrt{\lambda_{t+j-1}}b_{j\,t+j-1\, st}=c_{sit}.
$$
According to the unitarity of the transformation $U$ and 
the condition $c_{ijk}c_{ijk}^\ast=1$,
one gets $\lambda_i=1/N$, $i=1,...,N$, and the the state
(\ref{psi1}) is a maximally entangled one, which shows that
with a less than maximally entangled state it is
impossible to give a unitary transformation that fulfills perfect
teleportations.

For $N=2$, taking
$c_{111}=c_{211}=c_{121}=c_{112}=c_{212}=c_{122}=
-c_{221}=-c_{222}=1$ (this choice satifies the condition
$c_{ijk}c_{ijk}^\ast=1$), we have
$O_{11}=I$, $O_{12}=\sigma_x$, $O_{21}=\sigma_z$, $O_{22}=i \sigma_y$,
where $\sigma_{x,y,z}$ are Pauli matrices and $I$ is the $2\times 2$ 
identity matrix. The unitary transformation $U$ is then equal to the joint
actions of the controlled-not gate $C_{NOT}$ and the Walsh-Hadamard
transformation $H$, as defined e.g. in \cite{qc1,qc2}.
This recovers the usual protocol
for teleporting two level quantum states given in \cite{Bennett93}.

When $N=2^m$ for some $m\in\Nb$, a case discussed in \cite{luigi},
$\Psi_1$ can be rewritten as
$$
\Psi_1=\prod_{i=1}^{m}\vert EPR>_i=\prod_{i=1}^{m}\frac{1}{\sqrt{2}}
(\vert \uparrow\uparrow >+\vert \downarrow\downarrow>)_i,
$$
where $\vert EPR>=\frac{1}{\sqrt{2}}
(\vert \uparrow\uparrow >+\vert \downarrow\downarrow>)$, an
EPR pair of spin-$\frac{1}{2}$ particles, is the sum of
all spin up and down states
and $\vert EPR>_i$ stands for the $i$-th EPR
with the first (resp. second) attached to the 
Hilbert space $H_2$ (resp. $H_3$). 
Therefore instead of a fully entangled
state of two $N$-level qubits, we only need $m$ pairs of
entangled two-level qubits. This conforms with the
discussions in \cite{luigi}.

Generally, the dimension $N_2$ of $H_2$ can be greater than $N_1$. 
As long as one prepares the entangled state of two qubits in
the Hilbert spaces  $H_3$ and the
sub Hilbert space ${\cal H}_2\subset H_2$, with dim(${\cal H}_2)=N_1$, the
above results are still valid. The entangled qubits attached to the
Hilbert spaces $H_2$ and $H_3$
establish a quantum transportation ``tunnel", i.e. a way to teleport
a qubit on $H_1$ to $H_3$. To transport the qubit 
on $H_1$ to $H_3$, this tunnel should be constructed in such
a way that the
entangled state is prepared with suitable coefficients $a_{ij}$.
Moreover, this tunnel should be ``broad" enough to let the quantum
information
go through, in the sense that the dimension $N_2$ of the
Hilbert space $H_2$ should not be so small 
that there is no unitary transformation
satisfying equation (\ref{cond}) for any kind of entangled state $\Psi$.

We consider now some special cases of teleportations when some
components of the initial state are zero. Without  losing generality,
let $\alpha_i\neq 0$ for $i=1,...,n_1$, $n_1<N_1$,
and $\alpha_i=0$ for $i=n_1+1,...,N_1$
(We remark that for a given
$N_1$-dimensional vector it is always possible to make some of its
components
to be zero by changing the basis. But such a 
basis transformations depends of course on the
components of the given vector, hence for an unknown quantum state
this kind of transformation has no pratical use).
 
The initial state to be teleported under the above hypothesis
can be written as 
$$
\Psi_0=\sum_{i=1}^{n_1}\alpha_i e_i \,.
$$
We take the dimension of $H_2$ to be $N_2=n_1<N_1$.
The entangled state used to teleport $\Psi_0$ can be prepared in
the following way: 
\be\label{aij1}
a_{ij}=\left \{ \ba{l}
\displaystyle
\frac{1}{\sqrt{n_1}}\,\delta_{ij},~~~~~~~j=1,...,n_1\\[4mm]
0,~~~~~~~~~~~~~~~j=n_1+1,...,N_1
\ea\right.
\ee
for $i=1,2,...n_1$. From (\ref{cond1}) we get
the unitary transformation (\ref{u}) with
$$
b_{i\,t+i-1\, st}=\frac{c_{sit}}{\sqrt{N_1}}
$$ 
for $t, t+i-1~({\rm mod~n_1})=1,...,n_1$, $i,s=1,2,...,N_1$,
the other coefficients $b_{jist}$ in (\ref{cond1}) being zero.

An example is the teleportation of an EPR pair
$\Psi_0=\vert\Psi_{EPR}>=\alpha\vert 01>+\beta\vert
10>$, $\vert\alpha\vert^2 + \vert\beta\vert^2=1$, 
as discussed in \cite{gorbachev}.
In this case $N_1=4$. $\Psi_0$ can be written as
$\alpha e_3+\beta e_2\equiv \alpha e_1^\prime+\beta e_2^\prime$.
The dimension of the auxiliary Hilbert space $H_2$ is only needed to be
$n_1=2$. The entangled state is given by $\Psi_1=
\frac{1}{\sqrt{2}}(f_1\otimes g_1^\prime+f_2\otimes g_2^\prime)
=\frac{1}{\sqrt{2}}(f_1\otimes g_3+f_2\otimes g_2)
=\frac{1}{\sqrt{2}}(\vert 101>+\vert 010>)$.

Here as $n_1=N_1/2=2$, instead of (\ref{aij1}), we may alternatively
take $a_{ij}=\frac{1}{\sqrt{n_1}}\,\delta_{ij}$ for $j=n_1+1,...,N_1$ and
$a_{ij}=0$ for $j=1,...,n_1$. Then the entangled state becomes
$\Psi_1=\frac{1}{\sqrt{2}}(f_1\otimes g_1+f_2\otimes g_4)
=\frac{1}{\sqrt{2}}(\vert 000>+\vert 111>)$, which is called
a GHZ triplet consisting of three two-level qubits and can be realized
experimently \cite{8,9}. The unitary transformation is given by
$b_{i\,t+i-1\, st}=c_{sit}/\sqrt{N_1}$ 
for $t=1,...,n_1$, $t+i-1~({\rm mod~n_1})=n_1+1,...,N_1$, $i,s=1,2,...,N_1$,
and the other coefficients $b_{jist}$ in (\ref{cond1}) being zero. For a
suitable choice of the sign of $c_{sit}$,
this recovers the result in \cite{gorbachev}.

We have studied the general properties of 
teleportation for finite dimensional discrete quantum states.
The protocol we presented is for generally given entangled states with
$N_3=N_1$. If $N_3>N_1$, one can always take a subspace ${\cal
H}_3\subset H_3$ such that
dim$({\cal H}_3)=N_1$ and prepare the entangled state in the Hilbert spaces
$H_2$ and ${\cal H}_3$. Accordingly the initial state $\Psi_0$ 
will be sent to the subspace ${\cal H}_3$.


\bigskip
\medskip
\begin{thebibliography}{99}

\bibitem{Bennett93}C.H. Bennett, G. Brassard, C. Crepeau, R. Jozsa, 
A. Peres, and W.K. Wooters, Phys. Rev. Lett. {\bf 70}, 1895 (1993).

\bibitem{vaidman94}L.Vaidman, Phys. Rev. A {\bf 49} 1473 (1994).

\bibitem{Braunstein98}S.L. Braunstein and H.J. Kimble, Phys. Rev. Lett.
{\bf 80}, 869 (1998).

\bibitem{gorbachev}
V.N.Gorbachev and A.I.Trubilko, {\it Quantum teleportation of EPR pair 
by three-particle entanglement}, quant-ph/9906110, to appear in
JETP, v. 118, 4(10) (2000).

\bibitem{bose}
S. Bose  and  V. Vedral, Phys. Rev. A {\bf 61}, 040101(2000).

\bibitem{samuel}
S.L. Braunstein, G.M. D'Ariano, G.J. Milburn and
M.F. Sacchi, Phys. Rev. Lett. {\bf 84}, 3486(2000).

\bibitem{luigi}
L. Accardi and M. Ohya, {\it Teleportation of general quantum states},
quant-ph/9912087.

\bibitem{werner}
R. F. Werner, {\it All Teleportation and Dense Coding Schemes},
quant-ph/0003070.

\bibitem{galv}
E. Galv\~{a}o and L. Hardy, Phys. Rev. A {\bf 62}, 012309(2000).

\bibitem{IOS}  K. Inoue, M. Ohya and H. Suyari, Physica D
{\bf 120}(1998)117-124.

\bibitem{BBPSSW}  C.H. Bennett, G. Brassard, S. Popescu, B. Schumacher,
J.A. Smolin and W.K. Wootters, Phys. Rev. Lett. {\bf 76}(1996)722--725.

\bibitem{GC} D. Gottesman and I.L. Chuang, Nature {\bf 402}, 390-393 (1999).

\bibitem{qc1}
A. Steane, {\it Quantum computing}, Reports on Progress in 
Physics {\bf 61} (1998)117-173.

\bibitem{qc2}
E. Rieffel and W. Polak, {\it An introduction to quantum
computing for non-physicists}, quant-ph/9809016, to appear
in ACM computing surveys, 2000.

\bibitem{8}
D. Bouwmeester, J. Pan, M. Daniell, H. Weinfurter, A. Zeilinger,
Phys. Rev. Lett. {\bf 82}, 1345(1999).

\bibitem{9}
R.J. Nelson, D.G. Cory, S. Lloyd, {\it Experimental demonstration of
Greenberger-Horne-Zeilinger correlations using nuclear 
magnetic resonance}, quant-ph/9905028.

\end{thebibliography}
\end{document}